\documentclass[prd,preprintnumbers,nofootinbib,aps]{revtex4-1}
\usepackage{graphicx}% Include figure files
\usepackage{dcolumn}% Align table columns on decimal point
\usepackage{bm}% bold math
\usepackage{epsfig,amsmath}
\usepackage[usenames,dvipsnames]{color}
\usepackage[pagebackref=false, colorlinks=true]{hyperref}
\usepackage{appendix}
\definecolor{redish}{rgb}{0.7,0.2,0.0}  % color defined in (r=red,g=green,b=blue) model
\definecolor{bluish}{rgb}{0.2,0.5,0.8}

\hypersetup{linkcolor=redish,          % color of internal links
                  citecolor=blue,        % color of links to bibliography
                  filecolor=magenta,      % color of file links
                  urlcolor=bluish}          % color of the url

\DeclareFontFamily{U}{rsfs}{}         % Formal Script            %
\DeclareFontShape{U}{rsfs}{m}{n}{<5> rsfs5 <6><7> rsfs7          %
  <8><9><10><10.95><12><14.4><17.28><20.74><24.88> rsfs10}{}     %
\DeclareMathAlphabet{\mathfs}{U}{rsfs}{m}{n}                     %

\newcommand{\ba}{\nopagebreak[3]\begin{eqnarray}}
\newcommand{\ea}{\end{eqnarray}}

\newcommand{\f}{\frac}
\def \w{\wedge}
\def \o{\omega}
\def \n{\nabla}
\def \a{\alpha}
\def \b{\beta}
\def \t{\tilde}
\def \O{\Omega}

\def \e{\epsilon}
\def \p{\partial}
\def \d{\Delta}

\def \a{\alpha}

\def \th{\theta}

\begin{document}
\title{Strong gravity Lense-Thirring Precession in Kerr and Kerr-Taub-NUT
spacetimes}
\author{Chandrachur Chakraborty}
\email{chandrachur.chakraborty@saha.ac.in}
\affiliation{Saha Institute of Nuclear Physics, 
Kolkata 700064, India}
\author{Parthasarathi Majumdar}
\email{bhpartha@gmail.com}
\affiliation{Physics Department, Ramakrishna Mission Vivekananda 
University, Belur Math 711202, India}
\begin{abstract}
An exact expression derived in the literature 
for the rate of dragging of inertial frames
(Lense-Thirring (LT) precession) 
in a general stationary spacetime, is reviewed.
The exact LT precession 
frequencies for Kerr, Kerr-Taub-NUT
and Taub-NUT spacetimes are explicitly derived. 
Remarkably, in the case of the zero angular momentum Taub-NUT spacetime, the 
frame-dragging effect is shown {\it not} to vanish, when 
considered for spinning test gyroscopes. The result
becomes sharper for the case of vanishing ADM mass
of that spacetime. We clarify how our results are 
consistent with claims in the recent 
literature of null orbital plane precession 
for NUT spacetimes.   

\end{abstract}

\maketitle

\section{Introduction} 

Stationary spacetimes with angular momentum (rotation) are known 
to exhibit an effect called Lense-Thirring (LT) precession whereby 
locally inertial frames are dragged
along the rotating spacetime, making any test gyroscope in such spacetimes
{\it precess} with a certain frequency called the LT precession
frequency \cite{schiff}. This frequency has been shown to decay as the 
inverse cube of the
distance of the test gyroscope from the source \cite{lt} for large 
enough distances where curvature effects are small,
and known to be proportional to the angular momentum of the
source. Most earlier analyses of the LT effect \cite{jh} assume large distances
($r>>M,~ M$ is the ADM mass of the rotating spacetime due to a compact
object) for the test gyroscope. 

 Such weak field analyses lead to the standard result for LT precession
frequency in the weak field approximation, given by \cite{jh,li2}
\begin{equation}
\vec{\O}_{LT}=\f{1}{r^3}[3(\vec{J}.\hat{r})\hat{r}-\vec{J}]
\label{we1}
\end{equation}
where, $\hat{r}$ is the unit vector along $r$ direction. 
In a recent work reported in ref. \cite{ks}, an alternative 
approach based on solving the geodesic equations of the test gyroscope numerically,
{\it once again} within the weak gravitational field approximation, 
is used to compute the frame-dragging effect for galactic-centre black holes. 
In another very recent related work \cite{hl}, Hackman and L\"ammerzahl
have given an expression of LT precession valid up to {\it first order} in the Kerr
parameter $a$ for a general axisymmetric Pleba\'nski-Demia\'nski
spacetime. The LT precession rate 
has also been derived \cite{gk2,gk} through solving 
the geodesic equations for both Kerr and Kerr-de-Sitter spacetimes 
at the {\it polar orbit}. These results are not applicable for orbits
which lie in orbital planes other than the polar plane. We understand that 
observations of precession due to locally inertial frame-dragging 
have so far been possible only for spacetimes whose curvatures
are small enough; e.g., the LT precession in the earth's gravitational
field which was probed recently by the LAGEOS 
experiment \cite{nat} and also by Gravity 
Probe B \cite{grav}. Though there has been so far no
attempt to measure LT precession effects 
due to frame-dragging in {\it strong gravity regimes} \cite{sb},
observational prospects of LT precession in 
strong gravity situations have been discussed in \cite{sp}. 
The problem of accretion onto compact objects also stands 
to be influenced by strong gravity physics, especially by 
an understanding of LT precession under such conditions. 
A recent work by Stone and Loeb \cite{rr2} has estimated 
the effect of weak-field LT precession on accreting matter 
close to compact accreting 
objects. Modifications due to strong gravity LT precession 
to such situations are not without interest.

In this paper, we 
%review in brief the derivation 
%given in  \cite{ns,tk} of the LT precession rate in a strong 
%gravity regime in a `Copernican' frame, examining 
%its domain of validity. This formula is also applicable for all orbits (not only 
%for polar and equatorial orbits), located at 
%various distances and different angles. In this sense, it is
%rather more general than the result of \cite{mtw}. Also, the oft-quoted weak-field 
%result (\ref{we1}) (in a `Copernican' frame) for 
%the LT precession rate is readily obtained 
%from this general result, inserting the metric for the desired spacetime. 
present a detailed analysis of the frame dragging phenomenon in 
Kerr-Taub-NUT spacetime, where the NUT charge is an additional 
feature with interesting consequences. The Kerr and Taub-NUT 
spacetimes emerge as special cases of this analysis. Also, the oft-quoted weak-field 
result (\ref{we1}) (in a `Copernican' frame) for the 
LT precession rate is readily obtained from this 
general result, inserting the metric for the desired spacetime.  

The paper is accordingly organized as follows : in 
section II, we present a brief derivation of LT precession frequency 
in any stationary spacetime, following 
ref.s \cite{ns,tk}. 
%We  further discuss some 
%s. pecial features of the frame-dragging effect 
%in a {\it stationary} spacetime within a `Copernican' 
%frame. 
In section III, we derive the exact LT precession 
rates in Kerr-Taub-NUT and Taub-NUT
spacetimes.  This is followed by a detailed 
discussion of the Taub-NUT spacetime and how 
one observes a non-vanishing LT precession despite a 
vanishing Kerr parameter, provided one looks at spinning
test gyroscopes. A possible analytic extension of the
Taub-NUT spacetime is also considered, to delineate 
the properties of the `horizon'. The rate of the LT precession
in Kerr spacetime is next derived in section IV, without 
invoking either the weak gravity approximation
or an approximation involving the Kerr parameter. The weak-field approximation 
is then shown to emerge straightforwardly 
from our general formulation. We end in section V with a 
summary and a discussion on future outlook.

\section{Derivation of Lense-Thirring precession frequency}

Let us consider an observer at rest in a stationary 
spacetime with a timelike Killing field $K$. 
The observer moves along an integral curve $\gamma (\tau)$ of $K$. 
So, her 4-velocity can be written as
\begin{equation}
 u=(-K^2)^{-\f{1}{2}}K
\label{s1}
\end{equation}

We can now choose an orthonormal tetrad ${e_\a}$ along $\gamma$ 
which is Lie-transported.
\begin{equation}
 L_K e_{\a}=0
\label{s2}
\end{equation}
where $\a=0,1,2,3$. As $e_0$ is just $u=\dot \gamma$ 
(where,`dot' denotes the differentiation with respect to $\tau$),
 $u$ is perpendicular to $e_1,e_2,e_3$ axes. We also have 
\begin{equation}
 <e_\a,e_\b>=\eta_{\a\b},
\end{equation}
where, $<,>$ this symbol implies the scalar product and
$\eta_{\a\b}=diag(-1,1,1,1)$. We can interpret $e_{\a}$ as axes at rest. 
 This choice is what is sometimes known as the `Copernican' frame.

We know that the spin of the gyroscope precesses with respect 
to that axes of rest and we are interested in the change 
of the spin relative to this system. We know that torsion
\begin{equation}
 T(K,e_i)=\n_K e_i-\n_{e_i}K-[K,e_i]=0
\label{s3}
\end{equation}
We know
\begin{equation}
 \o_{ij}=<\n_ue_i, e_j>
\label{s3a}
\end{equation}
in where the $\o_{ij}$ related with the angular velocity $\O^l$ as
\begin{equation}
\o_{ij}=\e_{ijl} \O^l
\label{s5}
\end{equation}
Now, using eqn. (\ref{s3a}) and eqn. (\ref{s1}) we get,
\begin{equation}
 \o_{ij}=(-K^2)^{-\f{1}{2}}<e_j,\n_Ke_i>
\label{s4}
\end{equation}
The gyroscope precesses with the angular velocity $\O$ relative 
to the tetrad frame $e_{\a}$,
$\O$ is considered as the angular velocity or
the precession rate of the Lense-Thirring precession.
As $[K,e_i]=L_Ke_i=0$, we get from the eqn.(\ref{s3}) is
\begin{equation}
 \n_K e_i=\n_{e_i}K
\label{s6}
\end{equation}
Substituting this result in eqn.(\ref{s4}) we get,
\begin{equation}
 \o_{ij}=(-K^2)^{-\f{1}{2}}<e_j,\n_{e_i}K>
=(-K^2)^{-\f{1}{2}}\n \t{K}(e_j, e_i)
\label{s7}
\end{equation}
where, $\t{K}$ is the one-form of $K$. 
Eqn.(\ref{s7}) reduces to (as $\o_{ij}$ is anti-symmetric)
\begin{eqnarray}
 \o_{ij} &=& -(-K^2)^{-\f{1}{2}}\f{1}{2}[\n \t{K}(e_i, e_j)-\n \t{K}(e_j, e_i)] 
\label{s8} \\
 &=& \f{1}{2}(-K^2)^{-\f{1}{2}} d\t{K}(e_i, e_j)
\label{s9}
\end{eqnarray}

So, the exact LT frequency of precession of test
gyroscopes in strongly curved stationary 
spacetimes, analyzed within a Copernican frame, is expressed 
as a co-vector given in terms of the timelike 
Killing vector fields $K$ of the stationary spacetime,
as (in the notation of ref. \cite{ns, tk})
\begin{eqnarray}
\t \O&=&\f{1}{2K^2}*(\t K \w d\t K) 
\label{fr}
\\
\text{or,}&&  \nonumber
\\
\O_{\mu}&=&\f{1}{2K^2} \eta_{\mu}^{~\nu \rho \sigma} K_{\nu}
\p_{\rho} K_{\sigma}~, ~\label{ltfr}
\end{eqnarray} 
where, $\eta^{\mu \nu \rho \sigma}$ represent the components of the
volume-form in spacetime
and $\t K$ \& $\t \O$ denote the one-form of $K$ \& $\O$, respectively. 
$\t \O$ will vanish for stationary spacetimes if and only if
 $(\t K \w d\t K)$ does. 

For the general stationary spacetime, we can use 
the coordinate basis form of $ K= \p_0$ and
the co-vector components are easily seen to be $K_{\mu} =
g_{\mu 0}$. This co-vector could also be written in the following form 
\begin{eqnarray}
\t K&=&g_{00} dx^0+g_{0i}dx^i 
\end{eqnarray}
The spatial components of the
precession rate (in the chosen frame) are
\begin{eqnarray}
\O_{k} = \f12 \f{\epsilon_{ijl}}{g_{00}\sqrt{-g}} \left[
  g_{0i,j} \left( g_{00}g_{kl} - g_{0k} g_{0l} \right) -g_{0i} g_{kl}
  g_{00,j} \right] ~. \label{oell}   
\end{eqnarray}

The vector field  corresponding to the LT precession 
co-vector in (\ref{oell}) can be
expressed as   
\begin{eqnarray}
\O = \f{1}{2} \f{\e_{ijl}}{\sqrt {-g}} \left[g_{0i,j}\left(\p_l - 
\f{g_{0l}}{g_{00}}\p_0\right) -\f{g_{0i}}{g_{00}} g_{00,j}\p_l\right]
\label{s25}
\end{eqnarray}

The remarkable feature of the above equation (\ref{s25}) is that it is 
applicable to any arbitrary stationary spacetime which is non-static; 
it gives us the exact rate of LT
precession in such a spacetime. For instance, a Taub-Newman-Unti-Tamburino
(NUT) {\cite{taub}-\cite{nut}} spacetime with vanishing 
ADM mass is known to be non-rotating, but still has an angular momentum (dual or `magnetic' mass
\cite{rs}); we use  eqn.(\ref{s25}) to compute the LT precession frequency in this case as well.

{%\color{red}It could be easily seen from eqn.(\ref{s25}) 

\section{Lense-Thirring precession in Kerr-Taub-NUT spacetime}

\subsection{The Kerr-Taub-NUT spacetime $a \neq 0~,~n \neq 0$}

The Kerr-Taub-NUT spacetime is geometrically a stationary, axisymmetric
vacuum solution of Einstein equation with Kerr parameter $(a)$ and
NUT charge $(n)$. If the NUT charge vanishes, the solution reduces to the
Kerr geometry. The metric of the Kerr-Taub-NUT spacetime is\cite{ml}
\begin{equation}
ds^2=-\f{\d}{p^2}(dt-A d\phi)^2+\f{p^2}{\d}dr^2+p^2 d\th^2
+\f{1}{p^2}\sin^2\th(adt-Bd\phi)^2
\label{lnelmnt}
\end{equation}
With 
\begin{eqnarray}\nonumber
\d&=&r^2-2Mr+a^2-n^2,  p^2=r^2+(n+a\cos\th)^2,
\\
A&=&a \sin^2\th-2n\cos\th, B=r^2+a^2+n^2.
\end{eqnarray}
As the spacetime has an intrinsic angular momentum (due to Kerr parameter $a$),
we can expect a non-zero frame-dragging effect. We get from eqn.
(\ref{s25}), the LT precession rate in Kerr-Taub-NUT spacetime
is 
\begin{eqnarray}
\vec{\O}_{LT}^{KTN}=\f{\sqrt{\d}}{p}\left[\f{a \cos\th}{p^2-2Mr-n^2}
-\f{a \cos\th-n}{p^2}\right]\hat{r}
+\f{a \sin\th}{p} \left[\f{r-M}{p^2-2Mr-n^2}
-\f{r}{p^2}\right]\hat{\th}
\label{kt}
\end{eqnarray}
In contrast to the Kerr spacetime, where the source 
of the LT precession is the Kerr
parameter (specific angular momentum) $a$, the Kerr-Taub-NUT
spacetime has an extra 
somewhat surprising feature : the LT precession does not
vanish even for vanishing Kerr 
parameter $a=0$, so long as the NUT charge $n \neq 0$. 
This means that though the orbital angular momentum $(J)$ 
of this spacetime vanishes, the spacetime does indeed exhibit 
an  intrinsic {\it spinlike} angular momentum (at the 
classical level itself) which we discuss below in more detail. One can show that 
inertial frames are dragged along this orbitally non-rotating 
spacetime with the precession rate
\begin{eqnarray}
\vec{\O}^{TN}_{LT}=\f{n\sqrt{\d|_{a=0}}}{p^3}\hat{r}
\label{tn1}
\end{eqnarray}
where, $p^2=r^2+n^2$.
Notice that the precession rate is independent of $\th$ and also that it
vanishes when the NUT charge vanishes, as already 
alluded to above. In fact, for $a=n=0$, the Kerr-Taub-NUT spacetime
reduces to the {\it static} Schwarzschild spacetime 
which of course does not cause any inertial frame dragging.
 We consider this curious phenomenon in somewhat more detail 
in the next subsection.

\subsection{The Taub-NUT spacetime $a=0$}

The Taub-NUT spacetime is geometrically a stationary,
non-rotating vacuum solution of 
Einstein equation with NUT charge ($n$).
The Einstein-Hilbert action requires no modification
to accommodate this NUT charge or ``dual mass" 
which is perhaps an intrinsic
feature of general relativity, being a gravitational 
analogue of a magnetic monopole in electrodynamics \cite{lnbl}. 

Consider the line element (of NUT spacetime), which is presented by
Newman et. al.\cite{msnr}
\begin{eqnarray}
 ds^2=-f(r)\left[dt+4n \sin^2\f{\th}{2}d\phi\right]^2+\f{1}{f(r)}dr^2
+ (r^2+n^2)(d\th^2+\sin^2\th d\phi^2)
\label{nut}
\end{eqnarray}
where,
\begin{equation}
 f(r)=\f{r^2-2Mr-n^2}{r^2+n^2}
\end{equation}
Here, $M$ represents the ``gravitoelectric mass" or `mass' and $n$ represents
the ``gravitomagnetic mass'' or `dual' (or `magnetic') 
mass of this spacetime. It is obvious that the spacetime 
(\ref{nut}) is not invariant under time reversal
$t \rightarrow -t$, signifying that 
it must have a sort of `rotational sense', once again 
analogous to a magnetic monopole in electrodynamics. One 
is thus led to the conclusion that the source of the 
nonvanishing LT precession is this  ``rotational sense'' 
arising from a nonvanishing NUT charge.
Without the NUT charge, the spacetime is clearly hypersurface 
orthogonal and
frame-dragging effects vanish, as already mentioned above. 
This {\it `dual' mass} has been investigated in detail in ref. 
\cite{sen2}, who also refer to it as an 
{\it `angular momentum monopole} \cite{rs} in Taub-NUT spacetime.

In the Schwarzschild coordinate system, $f(r)=0$ at
\begin{equation}
r=r_{\pm}=M\pm\sqrt{M^2+n^2}
\end{equation}
$r_{\pm}$ are similar to {\it horizons} in this geometry 
in the sense that $f(r)$ changes sign from positive to 
negative across the horizon and the radial coordinate 
$r$ changes from spacelike to timelike. But is $r=r_+$ 
an event horizon in the sense of the event horizon of 
Schwarzschild spacetime ? We shall focus on this issue 
momentarily. For the present, we note that the LT 
precession rate (which can be easily obtained from 
eqn.(\ref{tn1}) also) is given by

\begin{equation}
 \vec{\O}^{MTN}_{LT}=\f{n(r^2-2Mr-n^2)^{\f{1}{2}}}{(r^2+n^2)^{\f{3}{2}}}\hat{r}
\label{tn2}
\end{equation}
It is clear that $\O^{MTN}_{LT}=0$ on $r=r_{\pm}$, 
in contrast to the LT precession frequency in the standard 
Kerr spacetime which is maximum closest to the event 
horizon ! Further, if we plot the magnitude of the 
precession rate as a function of the radial coordinate 
for $r > r_+$, as obtained from (\ref{tn2}), 
one obtains the profile like FIG.1.

\begin{figure}
    \begin{center}
\includegraphics[width=3in]{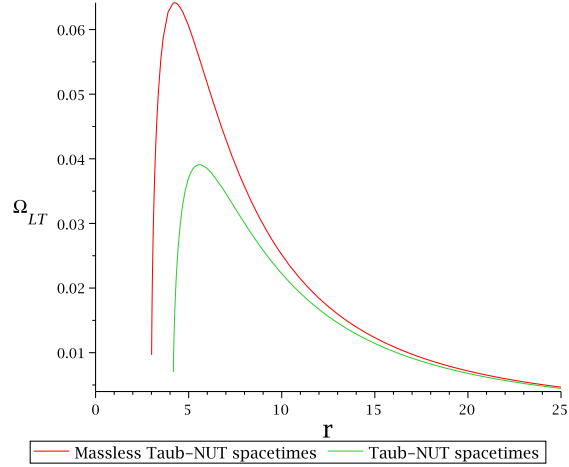}
      \caption{Plot of $\O_{LT}$ (in $m^{-1}$) vs $r$ (in $m$)
 for $n=3\,\, m$ \& $M=1\,\, m$ and $\O_{LT}$  vs $r$ for $n=3\,\, m$ \& $M=0$}
      \end{center}
\label{fig_2}
\end{figure}
Thus, the precession rate is maximum around $r=5$, but it sharply drops for 
$r \rightarrow r_+$ 
% most likely becoming ill-defined on the `horizon'. 
and vanishes on the `horizon'.

\subsection{Analytic extension of Taub-NUT spacetime}

As the metric (\ref{nut}) blows up at $r=r_{\pm}$, 
we should perhaps try a different co-ordinate system 
where it is smooth on the `horizon'. Following \cite{mkg}, 
wherein an analytic extension of the metric (\ref{nut}) 
has been attempted, one obtains the transformed metric
\begin{eqnarray}\nonumber
 ds^2&=&(r^2+n^2)(d\theta^2+\sin^2\theta d\phi^2)
\\
&+&F^2\left[du_{\pm}^2-dv_{\pm}^2-(2n/r_{\pm})(u_{\pm}dv_{\pm}-
v_{\pm}du_{\pm})\cos \theta d\phi-(n/r_{\pm})^2(u_{\pm}^2-v_{\pm}^2)
\cos^2\theta d\phi^2\right]
\end{eqnarray}
where,
\begin{eqnarray}
 F^2=4r_{\pm}^4(r^2+n^2)^{-1}\left(\f{r-r_{\mp}}{r_{\pm}}\right)^{1-\f{r_{\mp}}{r_{\pm}}}
\exp\left(-\f{r}{r_{\pm}}\right)
\\
u_{\pm}=\left(\f{r-r_{\pm}}{r_{\pm}}\right)^{1/2}\left(\f{r-r_{\mp}}
{r_{\pm}}\right)^{\f{r_{\mp}}{2r_{\pm}}}
\exp \left(\f{r}{2r_{\pm}}\right)\cosh \left(\f{t}{2r_{\pm}}\right)
\\
v_{\pm}=\left(\f{r-r_{\pm}}{r_{\pm}}\right)^{1/2}\left(\f{r-r_{\mp}}
{r_{\pm}}\right)^{\f{r_{\mp}}{2r_{\pm}}}
\exp \left(\f{r}{2r_{\pm}}\right)\sinh \left(\f{t}{2r_{\pm}}\right)
\end{eqnarray}
In $u, v$ co-ordinate system $r$ could be redefined as
\begin{equation}
 u_{\pm}^2-v_{\pm}^2=\left(\f{r-r_{\pm}}{r_{\pm}}\right)\left(\f{r-r_{\mp}}
{r_{\pm}}\right)^{\f{r_{\mp}}{r_{\pm}}}
\exp \left(\f{r}{r_{\pm}}\right)
\end{equation}

Recall that locally every spherically symmetric four 
dimensional spacetime has the structure 
${\cal I}_2 \otimes S^2$ where ${\cal I}_2$ is a two 
dimensional Lorentzian spacetime. In this 
Taub-NUT case, the attempted analytic extension discussed 
immediately above leads to a {\it vanishing} of the two 
dimensional Lorentzian metric on the `horizon' $r=r_+$, 
in contrast to the Schwarzschild metric. 
This might be taken to imply that perhaps the null 
surface $r=r_+$ is {\it not quite} an event horizon; 
rather it is a null surface where ingoing future-directed
 null geodesics appear to {\it terminate}, as already 
noticed in \cite{kag}. So, physical effects on this null hypersurface might not 
be easy to compute, as a result of which the apparent 
vanishing of the LT precession on this hypersurface is 
to be taken with a pinch of salt.   

The NUT spacetime, for the mass $M=0$ is also well-defined
(see, for example, appendix of \cite{rs}). We can also write down the precession
rate only for massless dual mass (NUT charge $n$ can be regarded as dual mass)
solutions of NUT spacetime. This turns out to be
\begin{equation}
 \vec{\O}^{TN}_{LT}=\f{n(r^2-n^2)^{\f{1}{2}}}{(r^2+n^2)^{\f{3}{2}}}\hat{r}
\label{tn3}
\end{equation}
% One may plot the precession frequency as a function of the 
% radial coordinate as earlier :
% \begin{figure}
% \begin{center}
% \includegraphics[width=5cm]{NUT.eps}
% \caption{Plot of $\O_{LT}^{TN}$  vs $r$ for $n=3\,\, m$ \& $M=0$}
% \end{center}
% \label{fig_2}
% \end{figure}
At, the points $r=\pm n$, the LT precession vanishes akin 
to the previous case, but the same caveats apply here as well.
 In FIG.1., we observe that for $n=3$, the LT 
precession starts for $r>3$
and continues to infinity. Setting 
$\f{d\O^{TN}_{LT}}{dr}=0$, we get that $\O^{TN}_{LT}$
is maximum at $r=\sqrt{2}n$. In our FIG.1., 
this value is $r=3\sqrt{2}=4.24$ m.
Now, we are not interested for $r<3$. Our 
formulas are not comfortable in that regions
and $r<r_{\pm}$ is also not well-defined for Taub-NUT spacetimes.
From our precession rate formulas 
(\ref{tn1},\ref{tn2},\ref{tn3}) at
dual mass spacetimes we can see that 
the precession rate $(\O^{TN}_{LT})$ is the same,
starting from the polar region to the equatorial 
plane for a fixed distance.
$\O^{TN}_{LT}$ depends only on distance $(r)$ of 
the test gyroscope from the `dual mass'.

At this point we refer to a recent paper by V. Kagramanova 
et al. \cite{kag} where it is claimed that the Lense-Thirring
 effect in fact vanishes {\it everywhere/everywhen} in the 
Taub-NUT spacetime. In that paper, timelike geodesic 
equations in this spacetime are investigated. The 
{\it orbital plane precession frequency}
$(\Omega_{\phi}-\Omega_{\theta})$ is computed, 
following the earlier work of ref. \cite{drasco, fh}, 
and a vanishing result ensues. This result is then 
interpreted in ref. \cite{kag} as a signature for a 
null LT precession in the Taub-NUT spacetime.

We would like to submit that what we have focused 
on in this paper is quite different from the 
`orbital plane precession' considered in \cite{kag}.
 Using a `Copernican' frame, we calculate the precession of a gyroscope
which is moving in an arbitrary integral 
curve (not necessarily geodesic). Within this 
frame, an untorqued gyro in a stationary but not 
static spacetime held fixed
by a support force applied to its center of mass, 
undergoes LT precession. Since the
Copernican frame does not rotate (by construction) relative to the
inertial frames at asymptotic infinity (``fixed stars"), the observed
precession rate in the Copernican frame also gives the precession rate
of the gyro relative to the fixed stars. It is 
thus, more an intrinsic property of the classical 
{\it spin} of the spacetime (as an untorqued gyro
 must necessarily possess), in the sense of a dual 
mass, rather than an {\it orbital} plane precession 
effect for timelike geodesics in a Taub-NUT spacetime. 
The dual mass is like the {\it Saha spin} of a magnetic 
monopole in electrodynamics \cite{lnbl}, which may have 
a vanishing orbital angular momentum, but to which a 
spinning electron must respond in that its wavefunction 
acquires a geometric phase.  

More specifically, in our case, we consider the 
gyroscope equation \cite{ns} in an 
arbitrary integral curve
\begin{equation}
 \nabla_u S=<S,a>u
\end{equation}
where, $a=\nabla_u u$ is the acceleration,
$u$ is the four velocity and $S$ indicates the spacelike classical spin 
four vector $S^{\alpha}=(0,\vec{S})$ of the gyroscope.
 For geodesics $a=0 \Rightarrow \nabla_u S = 0$.

In contrast,  Kagramanova et al. \cite{kag} consider the 
behaviour of massive test particles with {\it vanishing spin}
 $S=0$ \cite{val}, and compute the orbital plane precession 
rate for such particles, obtaining a vanishing result. 
We are thus led to conclude that because two different 
situations are being considered, there is no 
inconsistency between our results and theirs.

In summary, we have noted in this subsection several 
subtleties of computing the LT precession rate on 
and near the `horizon' of a Taub-NUT spacetime, 
and our results are consistent with earlier 
literature where geodesic incompleteness on this 
null hypersurface has been noted. 

\section{Lense-Thirring precession in Kerr spacetime}

One can now use eqn. (\ref{s25}) to calculate the angular momentum of
a test gyroscope in a Kerr spacetime to get the 
LT precession in a strong gravitational
field. In Boyer-Lindquist coordinates, the Kerr metric is written as,
\begin{eqnarray}
ds^2=-\left(1-\f{2Mr}{\rho^2}\right)dt^2-\f{4Mar \sin^2\theta}{\rho^2}d\phi dt
+\f{\rho^2}{\Delta}dr^2  
+\rho^2 d\theta^2+\left( r^2+a^2+\f{2Mra^2 \sin^2\theta}{\rho^2}\right) \sin^2\theta d\phi^2 
\label{k1}
\end{eqnarray}
where, $a$ is Kerr parameter, defined as $a=\f{J}{M}$, 
the angular momentum per unit mass and
\begin{equation}
 \rho^2=r^2+a^2 \cos^2\theta,        \\          \Delta=r^2-2Mr+a^2
\label{k2}
\end{equation}
For the Kerr spacetime, the only nonvanishing  $g_{0i} = g_{0
  \phi}$, $i = \phi$ and $j, l = r, \theta$; substituting these in
eq. (\ref{s25}), the precession frequency vector is given by 
\begin{eqnarray}
 \O_{LT}=\f{1}{2\sqrt {-g}} \left[\left(g_{0\phi,r}- \f{g_{0\phi}}{g_{00}} 
g_{00,r}\right)\p_{\theta}
 - \left(g_{0\phi,\theta}- \f{g_{0\phi}}{g_{00}} g_{00,\theta}\right)\p_r\right]
\label{k3}
\end{eqnarray}
where, the various metric components can be read off from 
eqn. (\ref{k1}). Likewise, 
\begin{equation}
\sqrt{-g}=\rho^2 \sin\theta
\label{k4}
\end{equation}

In order to make numerical predictions for the LT precession frequency
in a strong gravity domain, we need to transform the precession
frequency formula from the coordinate basis to the orthonormal
`Copernican' basis: first note that
 \begin{eqnarray}  
\O_{LT}&=& \O^{\theta} \p_{\theta}+\O^r \p_r \\ 
\O_{LT}^2&=&g_{rr}(\O^r)^2+g_{\theta\theta}(\O^{\theta})^2
\end{eqnarray}

Next, in the orthonormal `Copernican' basis at rest in the rotating
spacetime, the tetrad vector $e_0 = u$, the tangent vector along the
integral curve of the timelike Killing vector $K$. In this basis, with
our choice of polar coordinates, $\O_{LT}$ can be written as 
\begin{equation}
 \vec{\O}_{LT}=\sqrt{g_{rr}}\O^r \hat{r}+\sqrt{g_{\theta\theta}}\O^{\theta} \hat{\th}
\label{k9}
\end{equation}
where, $\hat{r}$ is the unit vector along the direction $r$. 
For the Kerr metric, 
\begin{equation}
 \O^{\theta}=-aM \sin\theta \f{(\rho^2-2r^2)}{\rho^4(\rho^2-2Mr)}
\label{k6}
\end{equation}
\begin{equation}
 \O^r=2aM \cos\theta \f{r\Delta}{\rho^4(\rho^2-2Mr)}
\label{k7}
\end{equation}
Substituting the values of $\O^r$ and $\O^{\theta}$ in
eqn.(\ref{k9}), we get the following expression of 
LT precession rate in Kerr spacetime 
\begin{eqnarray}
\vec{\O}_{LT}= 2aM \cos\theta \f{r\sqrt{\Delta}}{\rho^3 (\rho^2-2Mr)}\hat{r} 
-aM \sin\theta \f{\rho^2-2r^2}{\rho^3(\rho^2-2Mr)}\hat{\th}
\label{k10}
\end{eqnarray}
The magnitude of this vector is 
\begin{eqnarray}
\O_{LT}(r,\theta)=\f{aM}{\rho^3(\rho^2-2Mr)}\left[4\Delta r^2 \cos^2\theta 
 +(\rho^2-2r^2)^2 \sin^2\theta \right]^\f{1}{2} 
\label{k11}
\end{eqnarray}
This is the LT precession rate where no weak gravity approximation has
been made. It should therefore be applicable to any rotating spacetime
like rotating black hole etc.
\\ 
% \subsection{Weak field limit} 
% 
% For large distances, the Kerr metric 
% is approximated as a Schwarzschild metric with 
% the cross term $(g_{\phi t}d\phi dt)$, that is
% \begin{equation}
%  ds^2=ds_{Sch}^2-\f{4Ma \sin^2\theta}{r} d\phi dt
% \label{w1}
% \end{equation}
% and it has been shown that \cite{jh}
% \begin{equation}
%  \vec{\O}_{LT}=\f{1}{r^3}[3(\vec{J}.\hat{r})\hat{r}-\vec{J}]
% \label{w2}
% \end{equation} 
% where $J=aM$, the angular momentum of 
% the rotating spacetime. For a rotating
% compact object $J$ could be determined by 
% $J\simeq I\omega\simeq MR^2\omega$(following eqn. (14.23) of \cite{jh}
% and for perfect formulation of Moment of inertia $I$, see \cite{bh}).
% 
% For large distances and in weak gravitational fields (where we can take $r>M$ 
% and $r>a$), the 2nd term of eqn.(\ref{k10}) reduces to 
% \begin{eqnarray}
% \sqrt{g_{\theta\theta}}\O^{\theta} \hat{r}=
% -\f{J \sin\theta}{r^3(1+\f{a^2}{r^2} \cos^2 \theta)^\f{3}{2}} 
% \f{(-1+\f{a^2}{r^2} \cos^2 \theta)\hat{r}}
% {(1+\f{a^2}{r^2} \cos^2 \theta-\f{2M}{r})}
% \simeq \f{J \sin\theta}{r^3}\hat{r}
% \label{w3}
% \end{eqnarray}
% Similarly, from the 1st term of eqn.(\ref{k10}) one obtains the following,
% \begin{eqnarray}\nonumber
% \sqrt{g_{rr}}\O^r \hat{r}=\f{2J \cos\theta .r^2\sqrt{1+\f{a^2}{r^2}
% -\f{2M}{r}}}{r^5(1+\f{a^2}{r^2}\cos^2 \theta)^\f{3}{2}
% {\left(1+\f{a^2}{r^2} \cos^2 \theta-\f{2M}{r}\right)}}\hat{\th}
% \simeq \f{2J \cos\theta}{r^3}\hat{\th}
% \label{w4}
% \end{eqnarray}
In the weak-field limit ($r>>M$), eqn.(\ref{k10}) reduces to
\begin{equation}
 \vec{\O}_{LT}(r,\theta)=\f{J}{r^3} \left[2 \cos\th \hat{r}
+ \sin\th \hat{\th}\right]
\label{w5}
\end{equation}
where, $\theta$ is the colatitude. The resemblance of this equation with
eq. (\ref{we1}) is unmistakeable.

% The LT precession for a general stationary metric in the weak field
% limit may also be derived from (\ref{s25}). In this approximation,
% \begin{eqnarray}\nonumber
%  g_{00}\simeq-1, g_{ij}=\delta_{ij}, \f{g_{0l}}{g_{00}}<<1,
% \end{eqnarray}
% Under these conditions eqn.(\ref{s25}) reduces to,
% \begin{equation}
%  \O\simeq \f{1}{2}\e_{ijl}g_{0i,j}\partial_l
% \end{equation}
% As $e_l\simeq \partial_l$, a test gyroscope precesses in such a weak
% gravitational field with an angular velocity
%In the weak field limit and for a general axisymmetric metric, 
%eqn.(\ref{s25}) reduces further to the expression of the LT precession
%\begin{equation}
% \vec{\O}\simeq \f{1}{2} \vec{\nabla} \times \vec{g},
%\end{equation}
%where, $\vec{g}\equiv (g_{01},g_{02},g_{03})$. Here, 1,2,3 indicate the
%space components in that spacetime.} 

We can visualize the difference between strong 
and weak gravity LT precession through a graphical representation. 
\begin{figure}
    \begin{center}
\includegraphics[width=2.5in]{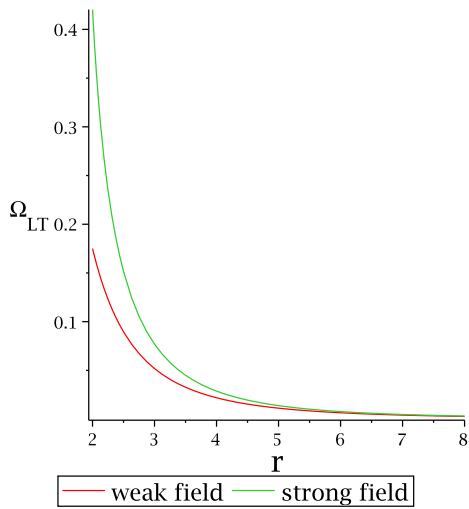}
      \caption{Plot of $\O_{LT}$ (in $m^{-1}$)  vs $r$ (in $m$)
 at $\th=0$ for $a=0.7\,\, m$ \& $M=1\,\, m$}
      \end{center}
\label{fig3}
\end{figure}
In FIG.2, we draw two graphs,
the red one is for $\O_{weakLT}=\f{2aM}{r^3}$ and the green 
 is for $\O_{strongLT}=\f{2aMr}{(r^2+a^2)^{\f{3}{2}} \sqrt{r^2-2Mr+a^2}}$
at $\th=0$. We see that $\O_{strongLT}$
is much much greater than $\O_{weakLT}$ for small $r$, i.e., 
near the compact body. As $r$ increases, the red and the green lines 
overlap with each other, i.e., the weak gravity approximation
emerges as a reasonable approximation.  

In a numerical comparison between
calculated values of $\O_{{\it strong}LT}$ and $\O_{{\it weak}LT}$ for typical
compact objects, with those of the sun and the earth, the
effect of strong gravity is seen to exceed by $30~\%$ the 
weak-gravity precession rate in general for all 
compact objects, while being roughly of the same order 
for the weak-gravity sources. This strongly motivates 
deeper observational probes of strong gravity LT precession of compact objects.

\section{Summary and Discussion}

The analyses presented above has two important features : (a) the LT precision 
frequency of a gyroscope in a `Copernican' frame within a Kerr spacetime 
is computed without any assumption on the angular momentum parameter or 
indeed the curvature of spacetime. The only comparable attempt in the 
literature is that in ref. \cite{mtw}, which however is not the 
same computation as ours, and the result is not the same in terms 
of metric coefficients. (b) The result derived in eqn. (\ref{s25}) 
is in fact valid, not just for axisymmetric spacetimes, but also for
 general non-static stationary spacetimes, once again without 
any assumptions about the curvatures involved. This result, we believe 
is applicable to a very large class of strong gravity systems. 

While most textbook calculations of the LT precession 
in the weak field approximation, the book of Misner, Thorne
and Wheeler, \cite{mtw} must also be mentioned. Here, 
the orbital angular velocity for {\it locally
non-rotating} observers in a Kerr-Newman spacetime is given in eqn.(33.24) 
as an exercise. This
formula does not appear to be restricted to the weak-field
approximation. However, from an astrophysical standpoint, it is not 
clear that the computed angular frequency corresponds to what might be 
measured as the Lense-Thirring precession in a strong gravity situation,
because it has been derived in a locally non-rotating frame which the 
authors amply clarify is {\it not} a {\it Copernican frame}.
A na\"ive limiting procedure does not appear to reduce this 
frequency to the standard weak-field result
(\ref{we1}) in `Copernican' frames quoted in most other textbooks for 
the LT precession rate in a weak gravitational field.

The substantial difference between the LT
precession frequency, arising 
in strong gravity regime and the standard, weak field precession
rate for inertial frame dragging 
ought to provide a strong motivation for their measurement
in space probes planned for the near future. The fascinating world of
gravitational effects associated with strongly gravitating compact
objects may provide the best yet dynamical observational signatures of
general relativity. In this paper, the focus has not been on 
understanding the effect of strong gravity LT 
precession on emission mechanism of pulsars and x-ray 
emission from black holes and neutron stars. We expect 
nontrivial modifications to arise from incorporation of 
frame-dragging effects in the theoretical analyses of 
these phenomena. We hope to report on this in the near future. There are 
other additional avenues of further work currently being 
explored : the most general axisymmetric solution of Einstein's 
equation given by the Pleba\'nski-Demia\'nski metric has been
investigated for an understanding of the LT precession 
in this case \cite{cp}. 

\noindent  {\it Note Added} : As suggested by an anonymous referee, 
incorporating multipole moment corrections (as example, for multipole corrections to the 
Schwarzschild metric see \cite{gw})
to the Kerr metric to determine the strong gravity LT precession rate near neutron
star surfaces is a very interesting project on which we hope to report elsewhere.
\\

\textbf{Acknowledgments :} We thank S. Bhattacharjee, R. Nandi,
A. Majhi, P. Pradhan and especially D. Bandyopadhyay for illuminating and helpful
discussions, and also for guiding us through the literature on
pulsar data. One of us (PM) thanks S. Bhattacharya for
discussions regarding observability of the LT precession in the strong
gravity regime. We also thank L. Iorio, C. L\"ammerzahl and N. Stone for  
useful correspondence regarding this paper and for bringing their work
relevant to the issues discussed here, to our attention. 
Last but not the least, one of us (CC) 
thanks V. Kagramanova and J. Kunz for gracious hospitality 
during an academic visit and for invaluable discussions 
regarding the subject of this paper. 
CC is also grateful to the Department of Atomic Energy
(DAE, Govt. of India) for the financial assistance.

\end{document}